\documentclass[twocolumn,showpacs,amsmath,amssymb,prb]{revtex4} 
\usepackage{graphicx}
\usepackage{dcolumn}
\usepackage{bm}
\usepackage{color}
\usepackage[colorlinks,bookmarks=false,citecolor=blue,linkcolor=blue,urlcolor=blue]{hyperref}
 
\begin{document} 
 
\title{Static impurities in the $S = 3/2$ kagome lattice} 
 
\author{A. L\"auchli} 
\affiliation{Institut Romand de Recherche Num\'erique en Physique des  
Materiaux (IRRMA), PPH--Ecublens, CH--1015 Lausanne, Switzerland}  
 
\author{S. Dommange}
\altaffiliation{deceased}
\affiliation{Institut Romand de Recherche Num\'erique en Physique des  
Materiaux (IRRMA), PPH--Ecublens, CH--1015 Lausanne, Switzerland}  
 
\author{B. Normand} 
\affiliation{D\'epartement de Physique, Universit\'e de Fribourg,  
CH--1700 Fribourg, Switzerland} 
\affiliation{Theoretische Physik, ETH--H\"ongerberg, CH--8093 Z\"urich,
Switzerland} 
 
\author{F. Mila} 
\affiliation{Institute of Theoretical Physics, Ecole Polytechnique 
F\'ed\'erale de Lausanne, CH--1015 Lausanne, Switzerland} 
 
\date{\today} 
 
\begin{abstract} 
 
We consider the effects of doping the $S$ = 3/2 kagome lattice with  
static, nonmagnetic impurities. By exact--diagonalization calculations 
on small clusters we deduce the local spin correlations and magnetization 
distribution around a vacancy. As in the $S$ = 1/2 kagome lattice, in the 
vicinity of the impurity we find an extended region where the spin 
correlations are altered as a consequence of frustation relief, and no 
indications for the formation of local moments. We discuss the implications 
of our results for local--probe measurements on $S$ = 3/2 kagome materials. 
 
\end{abstract} 
 
 
\maketitle 
 
\section{Introduction} 

The kagome antiferromagnet (Fig.~1) is one of the most highly frustrated 
geometries known in two--dimensional (2d) systems with only nearest--neighbor 
interactions, for both classical and quantum spins. A variety of materials 
displaying the kagome structure is known to exist, and their number 
continues to increase. Because of the long--standing absence of a true $S$ 
= 1/2 kagome spin system, the majority of experimental studies of quantum 
kagome antiferromagnets has focused on compounds with higher spins. These 
include the jarosites (H$_3$O)Fe$_3$(OH)$_6$(SO$_4$)$_2$,\cite{rwhrs} and 
KFe$_3$(OH)$_6$(SO$_4$)$_2$, both $S$ = 5/2, and 
KCr$_3$(OH)$_6$(SO$_4$)$_2$\cite{rkea,rlea}, which has $S$ = 3/2. In the 
magnetoplombite SrCr$_{9p}$Ga$_{12-9p}$O$_{19}$ (SCGO),\cite{rolitrp} 
most of the $S$ = 3/2 Cr$^{3+}$ ions form kagome bilayer units with 
frustrated interlayer coupling. The related compound 
Ba$_2$Sn$_2$ZnCr$_{7p}$Ga$_{10-7p}$O$_{22}$ (BSZCGO)\cite{rhhgrc} 
contains the same units with a superior interbilayer separation and a 
lower intrinsic impurity concentration. In both of the latter materials, 
ideal stoichiometry ($p = 1$) remains unachievable, making the influence 
of static (and for Ga$^{3+}$ spinless) impurities an important factor 
determining the physical response. 

The very recent synthesis of the $S$ = 1/2 kagome compound 
ZnCu$_3$(OH)$_6$Cl$_2$\cite{rsnbn} makes a detailed comparison of these 
different systems indispensible. Among the local--probe techniques which 
have been refined for specific studies of impurities in spin systems, 
nuclear magnetic resonance (NMR)\cite{rlmcmombm,rbmcb,roknsbnba,rinbsn} 
and muon spin resonance ($\mu$SR)\cite{rbmcbbabh,roknsbnba} experiments 
have provided the most valuable information obtained to date. Inelastic 
neutron scattering measurements offer additional insight into the 
excitations of the bulk system, suggesting an exotic, gapless 
spin--liquid state for ZnCu$_3$(OH)$_6$Cl$_2$ ($S$ = 1/2)\cite{rhea} 
and a rather conventional spin--wave spectrum with a quasi--flat band 
for KFe$_3$(OH)$_6$(SO$_4$)$_2$ ($S$ = 5/2),\cite{rmgnyhlnl} where 
Dzyaloshinskii--Moriya interactions have a significant role. 

\begin{figure}[t!] 
\medskip
\includegraphics[width=0.6\linewidth]{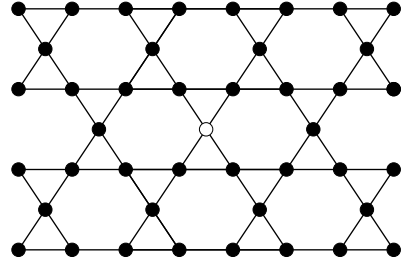}
\medskip
\caption{The kagome lattice. The central site has been replaced by 
a nonmagnetic impurity.} 
\label{fig:1} 
\end{figure} 
 
On the theoretical level, studies of the classical kagome antiferromagnet 
are complicated by the infinite ground--state degeneracy. For quantum 
spin systems, the $S$ = 1/2 kagome lattice has been shown to have a 
spin--liquid ground state,\cite{rce} with an exceptionally short spin--spin 
correlation length of 1.3 lattice constants.\cite{rle,rze} The excitation 
spectrum\cite{rlblps} is dominated by an exponentially large number of 
low--lying singlets.\cite{rm} Both ground and singlet excited states are 
found to be very well described by a resonating--valence--bond (RVB) 
framework based only on the manifold of nearest--neighbor dimer 
coverings.\cite{rmm}

By contrast, the situation for the $S$ = 3/2 system has been the subject of 
rather less attention. An early spin--wave analysis\cite{rc} suggested that 
it may support long--ranged antiferromagnetic order, but a large--$N$ 
approach\cite{rs} was found to favor a quantum--disordered state with 
the same properties (including a triplet gap and deconfined spinon 
excitations) as predicted for the $S$ = 1/2 system. Preliminary numerical 
results\cite{rldfm} for the $S$ = 3/2 system are complex and difficult to 
interpret, making it not yet possible to give an unambiguous statement on 
the underlying physics of this system: there is no specific evidence either 
for an ordered state or for a valence--bond description, and while the 
excitation spectrum shows certain parallels to the $S$ = 1/2 case, there 
are no indications for either spin waves or spinons.

Studies of static vacancies in spin liquids and gapped quantum magnets have 
been used in a number of systems to obtain additional information concerning 
spin correlations, and occasionally to reveal novel phenomena. The only 
investigations performed to date for kagome antiferromagnets have addressed 
the two limits, namely the maximally quantum and the purely classical. For 
the $S$ = 1/2 system it was shown in Ref.~\onlinecite{rdmnm} that nonmagnetic 
impurities do not generate free spin degrees of freedom in their vicinity, 
that they lower the number of low--lying singlet states, induce interdimer 
correlations over a significant range despite the very short spin correlation 
length, and experience a highly unconventional mutual repulsion; much of this 
exotic behavior is also contained within the RVB description. For the 
classical system, it was shown in Ref.~\onlinecite{rschb} that site disorder 
competes with thermal selection to favor states where frustration relief 
takes the form of noncoplanar spin configurations around the impurity site. 
Similar ideas were also articulated in Ref.~\onlinecite{rh}.

In this study we investigate the effects of static impurities in the 
$S$ = 3/2 kagome lattice with a view to offering a sound basis for the 
interpretation of experimental results. We will show that, as in the $S$ 
= 1/2 system, spin correlations are modified over a significant number of 
bonds at different distances from the impurity site, and that there is no 
evidence for free local moments induced around these sites. We provide a 
brief and qualitative motivation for these phenomena in terms of changes 
in the spin correlations analogous to the local collinearity enhancement 
of classical spins due to the relief of frustation at an impurity.

In Sec.~II we consider the spin--spin correlations on all bonds 
in the presence of an impurity in a $S$ = 3/2 kagome cluster. Section 
III presents the local magnetization distribution around an impurity 
site, which we compute for all spin sectors. In Sec.~IV we discuss
the extent to which our results may assist in the interpretation of 
NMR, $\mu$SR and other measurements on the $S$ = 3/2 kagome systems 
now under experimental investigation. Section V summarizes our 
conclusions. 
 
\section{Spin correlations}

\subsection{Exact Diagonalization}

We perform exact--diagonalization (ED) calculations for the Heisenberg 
Hamiltonian 
\begin{equation}
H = J \sum_{\langle ij \rangle} {\bf S}_i {\bf \cdot S}_j ,
\label{ehh}
\end{equation}
where $J$ is the antiferromagnetic superexchange interaction and $\langle 
ij \rangle$ denotes nearest--neighbor sites, for small clusters of $S$ = 
3/2 spins with periodic boundary conditions. Because of the large Hilbert 
spaces required when dealing with spins $S > 1/2$, the cluster size is 
restricted to a maximum of 15 sites with one impurity. We compute the 
total energy, the bond spin--spin correlation functions $C_{ij} = \langle 
{\bf S}_i {\bf \cdot S}_j \rangle$ for all nearest--neighbor pairs $\langle 
ij \rangle$ and the induced magnetizations on each site in every spin sector.

Specifically, we have studied standard kagome clusters of 12 and 15 sites 
in which one $S$ = 3/2 spin is replaced by a nonmagnetic impurity. In the 
12--site clusters all sites are equivalent before dilution, whereas in the 
15--site case there are two types of site which are inequivalent under 
the symmetries of the cluster. We have verified that our conclusions do 
not depend on choice of the dilution site. We note here that the ground 
state of the diluted 12--site system has total spin $S$ = 1/2, while the 
diluted 15--site cluster has a singlet ground state.

\begin{figure}[t!] 

\includegraphics[width=0.5\linewidth]{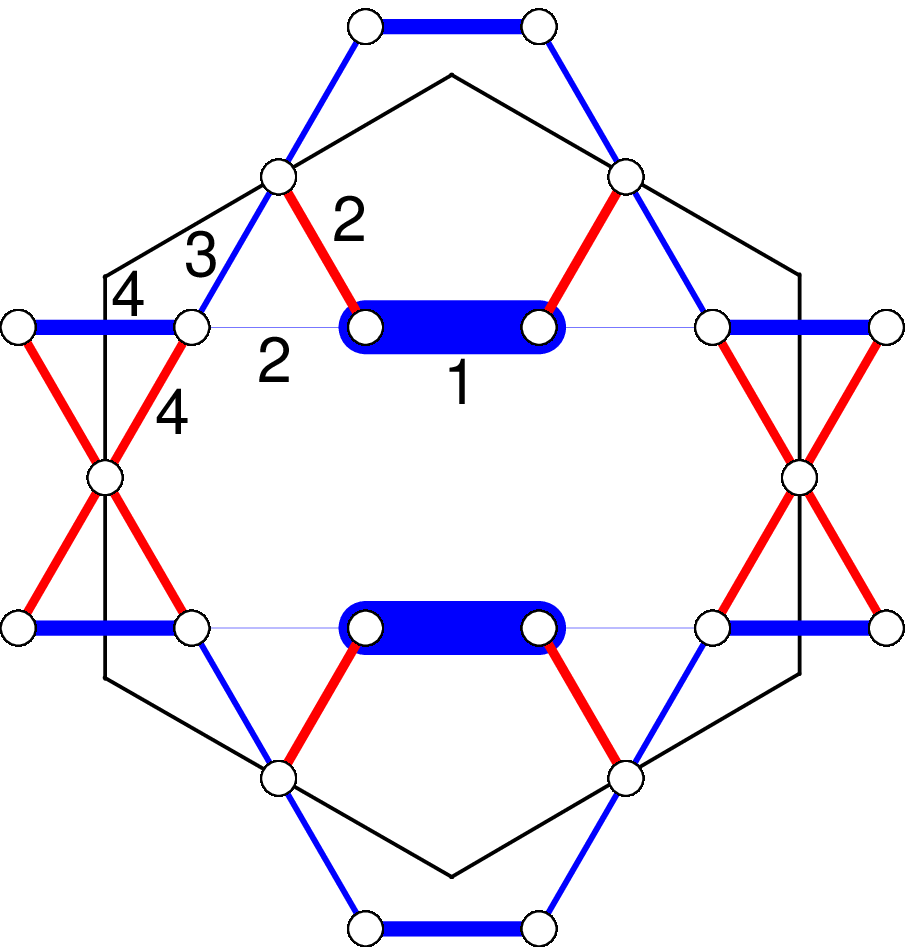}

\medskip

(a) $N = 12$

\medskip

\includegraphics[width=\linewidth]{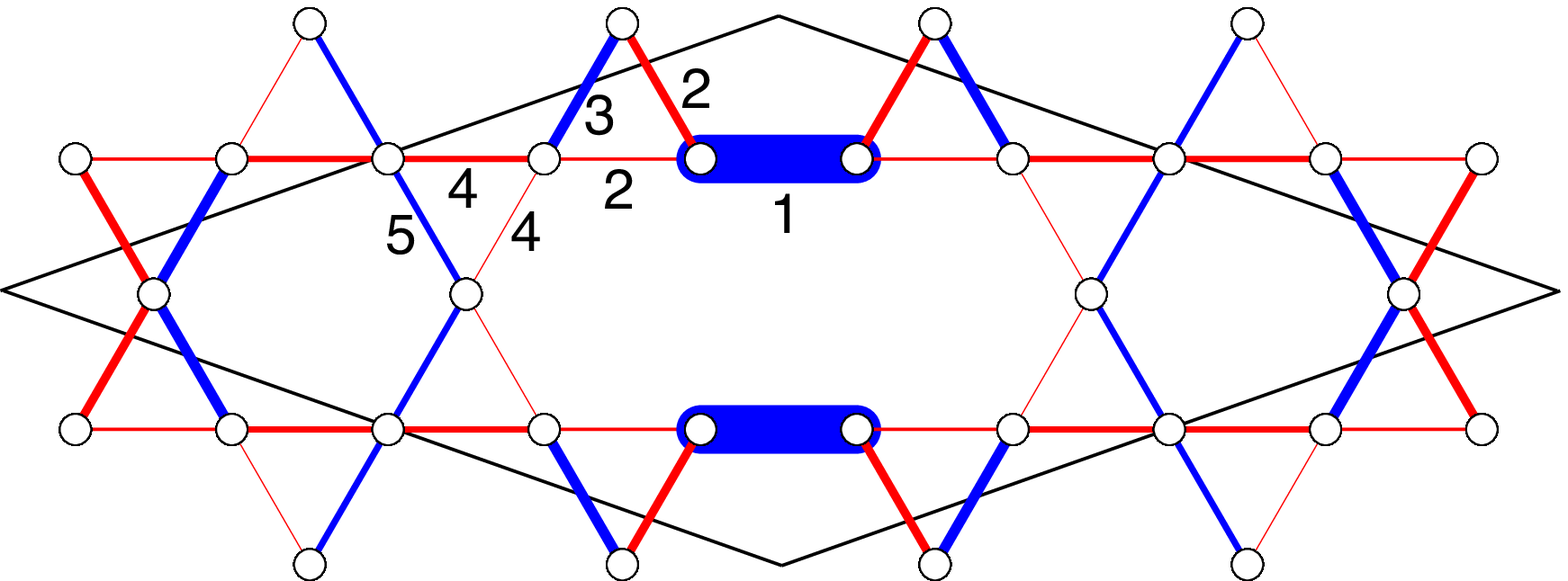}

\medskip

(b) $N = 15$

\caption{Bond spin--spin correlations for a single impurity in 12-- and 
15--site clusters, obtained by ED. Bar width represents the strength of 
correlation functions on each bond, measured as a deviation from the 
pure--system values $C_0 (12) = E_g(12) / J = -1.463$ and $C_0 (15) = 
E_g(15) / J = -1.448$, on a linear scale where the strongest correlation 
function is $\langle {\bf S}_i {\bf \cdot S}_j \rangle = -2.994$. Blue 
(red) lines denote bonds on which the deviation is negative (positive). 
The black lines denote the boundaries of the cluster.}
\label{fig:2} 
\end{figure} 

\subsection{Spin--spin correlation function}

The bond spin--spin correlation functions $C_{ij}$ around a single 
impurity are shown in Fig.~2. It is clear that spin correlations 
are strengthened on the bonds neighboring the vacant site. For a 
quantitative measure of this effect, we remind the reader that the 
spin correlation, or Heisenberg energy per bond $E = J {\bf S}_i \cdot 
{\bf  S}_j$, takes the values $E_0 = -15J/4$ if the two spins form a pure 
singlet, $E_1 = -11J/4$ for a triplet, $E_2 = -3J/4$ for a quintet, and 
$E_3 = 9J/4$ for a heptet state. The pure--system results $E_g(N = 12) = 
-1.463 J$ and $E_g(N = 15) = -1.448 J$ per bond are used to set the 
positive and negative values shown in Fig.~2, and also represent the 
average extent to which the bonds are dissatisfied (frustrated) in 
the $S$ = 3/2 kagome antiferromagnet. 

\begin{figure}[t!] 
\includegraphics[width=0.9\linewidth]{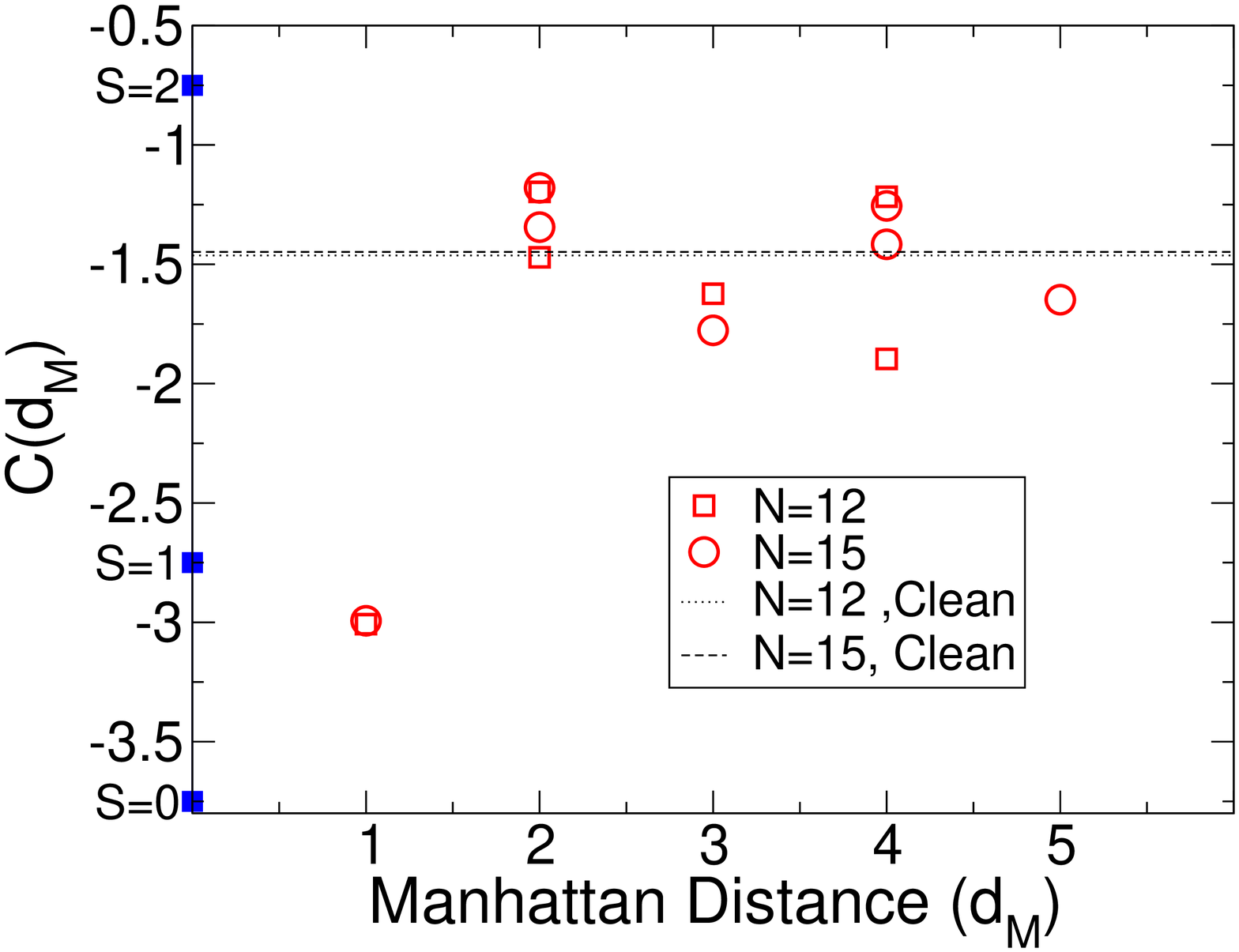}
\caption{Bond spin--spin correlation functions $C(d_{\rm M}) = \langle 
{\bf S}_i \cdot {\bf  S}_j \rangle$ on nearest--neighbor bonds as a 
function of their Manhattan distance $d_{\rm M}$ from a single impurity, 
for clusters of 12 and 15 sites. The site--bond separation $d_{\rm M}$ 
corresponds to the site labels shown in Fig.~\protect{\ref{fig:2}}. The 
dotted and dashed lines show the value of $\langle {\bf S}_i {\bf \cdot 
S}_j \rangle$ for impurity--free systems of sizes $N$ = 12 and $N$ = 15,
and are almost indistinguishable.}
\label{fig:3} 
\end{figure} 

The strengthening of correlations is manifestly not as strong as in the 
$S$ = 1/2 case, and the system remains far from perfect singlet formation 
on the nearest--neighbor bonds. For $S$=3/2, the correlation on
the strongest bond near the impurity (number 1 in Fig.\ref{fig:2}) represents
 80\% of the maximal (negative) possible value, and the next one 
(number 3 in Fig.\ref{fig:2}) 47\%  for the 15-site cluster, to be compared
with 92\% and 70\% for the $S$=1/2 case (taken from Ref.~\onlinecite{rdmnm}
for 27 sites). 

The most straightforward interpretation of the strengthening of correlations
is that the two spins on each bond attain a more antiferromagnetic 
state as a consequence of the relief of frustration in their triangle\cite{re} 
caused by the removal of the apical site; this effect was represented as an 
enhancement of collinearity for classical spins in Ref.~\onlinecite{rschb}.
A corollary of this enhancement is the weakening of spin correlations on 
next--nearest--neighbor bonds. A certain oscillatory behavior of the spin 
correlations is discernible in the values $C_{ij}$ as the separation of the 
bond from the impurity is increased, as shown in Fig.~3. Although this 
appears neither as strong nor as long--ranged as in the $S$ = 1/2 case, 
the effect clearly extends well beyond the nearest--neighbor sites. Unlike 
the $S$ = 1/2 case, these spin correlations do not illustrate directly the 
absence of local moments, which we demonstrate in Sec.~III. While the size 
of the cluster sets an obvious limit on the strength of these statements, 
we may conclude that impurity doping causes a disruption of the pure--system 
spin configuration which is not entirely local, and induces at least 
short--ranged dimer correlations with a characteristic length of several 
lattice constants.

\begin{figure}[t!] 
\includegraphics[width=\linewidth]{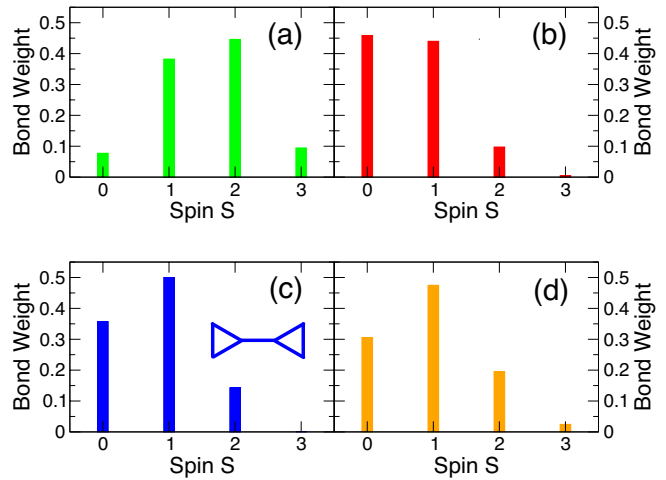}
\caption{Distribution of bond spin states obtained from ED calculations on 
the $S$ = 3/2 kagome antiferromagnet for (a) a bond far from an impurity, 
computed for an undoped 12--site cluster, (b) a bond next to an impurity, 
obtained from a 15--site cluster, and (c) the central bond of the 6--site 
cluster shown in the inset. In panel (d) is shown the analogous bond 
distribution for an undoped $S$ = 3/2 square lattice. }
\label{fig:4} 
\end{figure} 

\subsection{Bond spin correlations}

Further insight into the nature of the spin correlations and their 
alteration in the presence of impurities may be gleaned from the 
weights of the 4 possible spin states present on a given bond, which 
are shown in Figs.~4(a--d). For a bond far from an impurity site 
[Fig.~4(a)], one observes that this weight distribution is approximately 
symmetrical, with the majority of bonds in states of total spin 1 or 2, 
while the probability for a bond to be a perfect singlet or fully 
spin--polarized is small. Figure 4(b) shows the situation for a bond 
next to an impurity site, where there is a very significant shift of 
weight away from high--spin states, particularly $S$ = 2, to the net 
singlet state. This result quantifies the quantum analog of the 
classical collinearity enhancement, although as noted in the previous 
section this is by no means as overwhelmingly strong as in the $S$ = 1/2 
system. In Fig.~4(c) the same distribution is presented for the central 
bond of the 6--site cluster shown in the inset, which has $\langle 
{\bf S}_i {\bf \cdot S}_j \rangle = -2.821$. From the similarity to 
Fig.~4(b) one may conclude that the physical processes contributing to 
the spin correlations on the bond with relieved frustration are largely 
local in nature. 

For further comparison, in Fig.~4(d) we show the same distribution 
calculated for a square lattice of spins $S$ = 3/2 with no impurities. 
If this distribution is considered to be representative of the situation 
in an unfrustrated, collinear 2d system, a comparison with Fig.~4(a), 
where the highest weight is found for bonds of spin 2, shows the effects 
of the frustrated kagome geometry in driving the system away from a 
satisifed antiferromagnetic state. From the bond spin distributions 
(Fig.~4) one may also comment on the suitability of a description for 
the $S$ = 3/2 kagome system based on local singlet formation. While 
this type of framework was used with considerable success in the $S$ 
= 1/2 case, it is clear immediately from Fig.~4(a), where over 80\% 
of the bonds in the pure system have spin states $S$ = 1 and 2, that 
a basis of local bond singlets would not be expected to capture the 
dominant physics in this case. 

\begin{figure}[t!] 
\includegraphics[width=\linewidth]{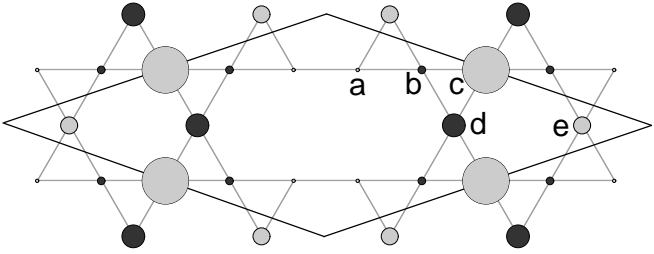}
\medskip
\caption{Site magnetization profile of the lowest triplet state induced 
by a single impurity in a 15--site cluster. Circle radius represents the 
magnitude of the local moment on a linear scale where the largest circle 
corresponds to a moment of 0.722$\mu_{\rm B}$ (0.5$\mu_{\rm B}$ is the 
moment of a free electron), and the dark grey circles represent sites 
with induced moments opposite to the effective field direction. The black 
lines denote the boundary of the cluster.} 
\label{fig:5} 
\end{figure} 

\section{Induced magnetization}

The local magnetization pattern induced in the vicinity of a doped  
nonmagnetic impurity is computed in the ED technique by considering for 
example the sector of total spin $(S,S_z)$ = (1,1), and is shown for a 
15--site cluster in Fig.~5. The site magnetizations are to some extent 
anticorrelated with the deviation of $C_{ij}$ from $C_0$ shown in Figs.~2 
and 3. The most obvious feature is the absence of local moments induced on 
the sites directly neighboring the impurity, a feature in common with the 
$S$ = 1/2 system but in contrast to the majority of gapped quantum magnets 
considered in this context. The induced moments are strongest on the sites 
most distant from the impurity, also suggesting, as in the $S$ = 1/2 case, 
a phenomenon of longer range where the moments would be found further from 
the impurity on a larger cluster. Such an extended magnetization profile 
is suggested by NMR and $\mu$SR measurements performed on SCGO and 
BSZCGO\cite{rlmcmombm,rbmcb,rbmcbbabh} (Sec.~IV).

The magnetization induced on each site by a nonmagnetic impurity is shown 
in Fig.~6 for all of the spin sectors obtainable on a 15--site cluster. The 
spin sector may be considered to reflect the state of the system under an 
applied magnetic field which changes the magnetisation from zero ($S_z = 0$) 
to saturation ($S_z = 21$). Unlike the $S$ = 1/2 case there is no apparent 
``protection'' of the nearest--neighbor sites (filled circles in Fig.~6), 
reflecting the fact that these are not bound into perfect dimers. The 
immediate polarization of the moment on these sites with the effective 
field is to be expected from the fact that the bond spin state is primarily 
a superposition of singlets and triplets [Fig.~4(b)]. The magnetization at 
these sites increases linearly to a local saturation at rather less than 
2/3 of the total saturation, as a consequence of their lower connectivity. 
The more distant sites show a complex and non--monotonic evolution of their 
local moments which suggests a rearrangement of the induced magnetization 
with applied field, and presumably strong effects of the finite cluster size in the 
calculation. 

\begin{figure}[t!] 
\includegraphics[width=0.9\linewidth]{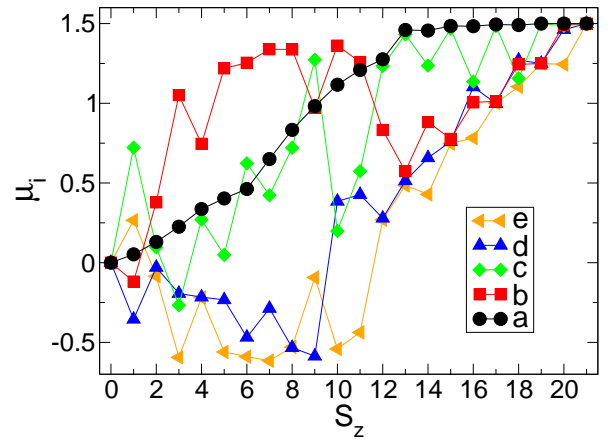}
\medskip
\caption{Induced magnetization per site $\mu_i$ as a function of the spin 
sector $S_z$ for all inequivalent sites neighboring an impurity, computed 
by ED for a 15--site cluster. The site labels correspond to those shown in 
Fig.~\protect{\ref{fig:5}}.} 
\label{fig:6} 
\end{figure} 

\section{Discussion}

NMR and $\mu$SR are the techniques of choice for measuring in real space  
the effects of doped impurities on magnetizations and spin correlations. 
In NMR it is the shifts, linewidth alterations and on occasion 
relaxation times which yield information concerning these quantities. In 
$\mu$SR it is the temperature--, field--, and doping--dependence of the 
muon relaxation time, the last being the most instructive with regard to 
the spatial extent of the spin excitations responsible for relaxation. 
The diluted $S$ = 3/2 kagome bilayer systems SCGO and BSZCGO have been 
the subject of an extensive program of investigation. It goes back to 
an early investigation of SCGO with $\mu$SR\cite{uemura} which opened
the way to more recent $\mu$SR investigations\cite{mendels}.  The most recent 
detailed results including both NMR and $\mu$SR  are contained 
in Refs.~\onlinecite{rlmcmombm}, 
\onlinecite{rbmcb}, and \onlinecite{rbmcbbabh}, and are summarized in 
Refs.~\onlinecite{rblmcb} and \onlinecite{mendels2}. Both materials have 
intrinsic doping of Cr$^{3+}$ 
sites by nonmagnetic Ga$^{3+}$ ions in the kagome bilayers, in quantities 
exceeding 5\% and 3\% respectively, although in BSZCGO there is evidence 
for a second type of defect state possibly related to Zn impurities. NMR 
studies including NQR (nuclear quadrupole resonance) measurements were 
performed on $^{71}$Ga sites in SCGO and BSZCGO, accompanied by 
$\mu$SR experiments on both materials. 

We abstract the primary features observed in experiment, and compare 
these to the results of the calculations presented in Secs.~II and III. 
We focus only on qualitative behavior, as a quantitative determination 
of any physical properties (shifts, line widths, relaxation rates) would 
require a specific Hamiltonian for local and hyperfine interactions. We 
caution the reader also that our calculations do not include the 
possibility of additional interactions breaking SU(2) spin symmetry on 
each bond; in particular the Dzyaloshinskii--Moriya interactions which 
may be present in these low--symmetry crystal structures are known to have 
a very significant effect on local magnetizations (Sec.~III)\cite{rkcm}.

Both experimental techniques find no evidence for a singlet--triplet gap 
at temperatures down to 30 mK, which is consistent with the presence of all 
types of states at low energies on a finite cluster. The line shape
observed in NMR shift measurements indicates the 
presence of many inequivalent sites, consistent with the large spatial extent 
of the region affected by a single impurity. The extended nature of the 
perturbation is also reflected in symmetric line shapes. The 
density of very low--energy spin excitations is manifest in $\mu$SR 
experiments as the observation of ``quantum dynamics'' persisting as 
$T \rightarrow 0$. The spatial extent of these collective (as opposed 
to local) spin fluctuations is reflected both in the doping--dependence 
of the relaxation rate and directly in the relaxation of muons at sites 
far from any impurities.

The important qualitative features of both the experiments and the 
numerical calculations are the absence of a single spin degree of freedom 
anywhere in the vicinity of the impurity, and the rather large extent of 
the area of affected sites in spite of the very short spin correlation 
lengths. We stress that these features, which are common to both $S$ = 
1/2\cite{rdmnm} and $S$ = 3/2, are common because of the highly frustrated 
kagome geometry: as noted in Secs.~I--III there is currently little 
evidence for a common description of the two systems. For the $S$ = 3/2 
Heisenberg model on the kagome lattice, a very small or vanishing spin 
gap and the near--degeneracy of considerable numbers of basis states are 
sufficient to yield the results we have obtained. There is as yet no 
framework providing an unambiguous understanding of the nature of the 
pure system, by which is meant providing ground and excited states 
satisfying these criteria, and it has been possible only to exclude  
some candidate descriptions, such as a discernible magnetic order or 
an origin in formation of local singlets. To date there is also no 
evidence for the possibility of deconfined, spinon--like excitations of 
the type sometimes invoked in the discussion of the $S$ = 1/2 system.

We reiterate that, as already shown for the $S$ = 1/2 case, kagome systems 
do not form part of the paradigm obeyed by many unfrustrated low--dimensional 
spin systems in which a local moment is formed near a doped vacancy. Instead, 
an extended envelope of weak magnetization is induced over a considerable 
range, and is maximal some distance from the impurity site. Despite the 
antiferromagnetic spin correlations, in kagome systems there is no sense 
in which the induced magnetization, or spin polarization, can be said to be 
staggered; this result is also common to the $S$ = 1/2 case,\cite{rdmnm} 
and is a consequence of the strongly frustrated geometry. An alternation is, 
however, found in the bond spin correlation functions, corresponding to an 
induced dimer--dimer correlation with a range of several lattice constants. 

\section{Summary}

We have investigated the effects of static, spinless impurities doped 
into the $S = 3/2$ antiferromagnetic Heisenberg model on the kagome 
lattice. By analysis of spin--spin correlation functions and of the 
induced local magnetization, we find that nonmagnetic impurities in the 
kagome lattice have the important property that they do not generate 
localized spins at the vacant site, and do have an extended range of 
influence on the correlated spin background. Unlike the $S$ = 1/2 case, 
where the development of dimer correlations over significant distances 
could be considered as due to a hole--induced freezing of singlet 
resonances,\cite{rdmnm} here it appears more appropriate to consider 
the phenomenon more generally as the result of enhanced local 
collinearity due to frustration relief. 

Our results are in excellent agreement with all of the qualitative 
features observed in local--probe experiments. This constitutes an 
important contribution to the definition of a set of generic properties 
of the doped kagome system. It also indicates that the materials SCGO 
and BSZCGO, despite their bilayer geometry, do share these kagome 
hallmarks. Because these systems do not have complete filling of the 
kagome lattice sites by $S$ = 3/2 ions, the study of impurity effects 
is a significant component of their characterization. 

We close by reminding the reader that the results we have presented are 
obtained for very small systems, and thus while indicative they cannot be 
considered as definitive. They should be regarded as provisional also in 
the sense that the ground state of and nature of spin correlations in the 
$S$ = 3/2 kagome antiferromagnet remain poorly characterized. However, 
they are sufficient to illustrate the general features which lie at the 
origin of the observed behavior in this system. 

\acknowledgments 
 
We are grateful to P. Mendels for valuable discussions. This work was 
supported by the Swiss National Science Foundation, both directly and 
through its MaNEP project.


\begin{thebibliography}{99} 
 
\bibitem{rwhrs} A. S. Wills, A. Harrison, C. Ritter, and  R. I. Smith, 
Phys. Rev. B {\bf 61}, 6156 (2000). 

\bibitem{rkea} A. Keren, K. Kojima, L. P. Le, G. M. Luke, W. D. Wu, 
Y. J. Uemura, M. Takano, H. Dabkowska, and M. J. P. Gingras, Phys. Rev. 
B {\bf 53}, 6451 (1996). 

\bibitem{rlea} S.--H. Lee, C. Broholm, M. F. Collins, L. Heller, 
A. P. Ramirez, Ch. Kloc, E. Bucher, R. W. Erwin, and N. Lacevic. 
Phys. Rev. B {\bf 56}, 8091 (1997). 

\bibitem{rolitrp} X. Obradors, A. Labarta, A.Isalgu\'e, J. Tejada, 
J. Rodriguez, and M. Pernet, Solid State Commun. {\bf 65}, 189 (1988).

\bibitem{rhhgrc} I. S. Hagemann, Q. Huang, X. P. A. Gao, A. P. Ramirez, 
and R. J. Cava, Phys. Rev. Lett. {\bf 86}, 894 (2001). 

\bibitem{rsnbn} M. P. Shores, E. A. Nytko, B. M. Barlett, and D. G. Nocera, 
J. Am. Chem. Soc. {\bf 127}, 13462 (2005).

\bibitem{rlmcmombm} L. Limot, P. Mendels, G. Collin, C. Mondelli, B.  
Ouladdiaf, H. Mutka, N. Blanchard, and M. Mekata, Phys. Rev. B {\bf 65},  
144447 (2002).
 
\bibitem{rbmcb} D. Bono, P. Mendels, G. Collin, and N. Blanchard, Phys. 
Rev. Lett. {\bf 92}, 217202 (2004). 

\bibitem{roknsbnba} O. Ofer, A. Keren, E. A. Nytko, M. P. Shores, 
B. M. Barlett, D. G. Nocera, C. Baines, and A. Amato, unpublished 
(cond--mat/0610540). 

\bibitem{rinbsn} T. Imai, E. A. Nytko, B. M. Barlett, M. P. Shores, and 
D. G. Nocera, unpublished (cond--mat/0703141).

\bibitem{rbmcbbabh} D. Bono, P. Mendels, G. Collin, N. Blanchard, F. Bert, 
A. Amato, C. Baines, and A. D. Hillier, Phys. Rev. Lett. {\bf 93}, 187201 
(2004). 

\bibitem{rhea} J. S. Helton, K. Matan, M. P. Shores, E. A. Nytko, 
B. M. Barlett, Y. Yoshida, Y. Takano, A. Suslov, Y. Qiu, J.--H. Chung, 
D. G. Nocera, and Y. S. Lee, Phys. Rev. Lett. {\bf 98}, 107204 (2007). 

\bibitem{rmgnyhlnl} K. Matan, D. Grohol. D. G. Nocera, T. Yildirim, 
A. B. Harris, S. H. Lee, S. Nagler, and Y. S. Lee, Phys. Rev. Lett. 
{\bf 96}, 247201 (2006). 

\bibitem{rce} J. T. Chalker and J. F. Eastmond, Phys. Rev. B {\bf 46}, 
14 201 (1992).  

\bibitem{rle} P. W. Leung and V. Elser, Phys. Rev. B {\bf 47}, 5459 (1993).  
 
\bibitem{rze} C. Zeng and V. Elser, Phys. Rev. B {\bf 51}, 8318 (1995).  

\bibitem{rlblps} P. Lecheminant, B. Bernu, C. Lhuillier, L. Pierre, and 
P. Sindzingre, Phys. Rev. B {\bf 56}, 2521 (1997); C. Waldtmann, H.--U. 
Everts, B. Bernu, C. Lhuillier, P. Sindzingre, P. Lecheminant, and 
L. Pierre, Eur. Phys. J. B {\bf 2}, 501 (1998).

\bibitem{rm} F. Mila, Phys. Rev. Lett. {\bf 81}, 2356 (1998). 
 
\bibitem{rmm} M. Mambrini and F. Mila, Eur. Phys. J. B {\bf 17}, 651 (2000).  

\bibitem{rc} A. Chubukov, Phys. Rev. Lett. {\bf 69}, 832 (1992). 

\bibitem{rs} S. Sachdev, Phys. Rev. B {\bf 45}, 12377 (1992).  

\bibitem{rldfm} A. L\"auchli, S. Dommange, J.--B. Fouet, and F. Mila, 
unpublished.

\bibitem{rdmnm} S. Dommange, M. Mambrini, B. Normand, and F. Mila, Phys. 
Rev. B {\bf 68}, 224416 (2003). 

\bibitem{rschb} E. F. Shender, V. B. Cherepanov, P. C. W. Holdsworth, 
and A. J. Berlinsky, Phys. Rev. Lett. {\bf 70}, 3812 (1993).

\bibitem{rh} C. L. Henley, Phys. Rev. Lett. {\bf 62}, 2056 (1989); 
Can. J. Phys. {\bf 79}, 1307 (2001).   

\bibitem{re} V. Elser, Phys. Rev. Lett. {\bf 62}, 2405 (1989).

\bibitem{uemura}   Y. J. Uemura, A. Keren, K. Kojima, L. P. Le, G. M. Luke, W. D. Wu, Y. Ajiro, T. Asan2, Y. Kuriyama, M. Mekata, H. Kikuchi, and K. Kakurai, Phys. Rev. Lett. {\bf 73}, 3306(1994).

\bibitem{mendels}  P. Mendels, A. Keren, L. Limot, M. Mekata, G. Collin, and M. Horvatic, 
Phys. Rev. Lett. {\bf 85}, 3496 (2000).

\bibitem{rblmcb} D. Bono, L. Limot, P. Mendels, G. Collin, and N. Blanchard, 
Low Temp. Phys. {\bf 31}, 704 (2005). 

\bibitem{mendels2} P. Mendels, A. Olariu, F. Bert, D. Bono, L. Limot, G. Collin, B. Ueland, 
P. Schiffer, R. J. Cava, N. Blanchard, F. Duc, J. C. Trombe, J. Phys.: Condens. Matter {\bf 19}, 145224 (2007).

\bibitem{rkcm} K. Kodama, S. Miyahara, M. Takigawa, M. Horvatic, 
C. Berthier, F. Mila, H. Kageyama, and Y. Ueda, J. Phys. Condens. Matter 
{\bf 17}, 61 (2005); M. Clemancy, H. Mayaffre, C. Berthier, M. Horvatic, 
J.--B. Fouet, S. Miyahara, F. Mila, B. Chiari, and O. Piovesana, Phys. 
Rev. Lett. {\bf 97}, 167204 (2006); S. Miyahara, J.--B. Fouet, S. Manmana, 
R. Noack, H. Mayaffre, I. Sheikin, C. Berthier, and F. Mila, Phys. Rev. B 
{\bf 75}, 184402 (2007).

\end{thebibliography}
\end{document}